\documentclass[a4paper,fleqn,usenatbib]{mnras}

\usepackage{newtxtext,newtxmath}

\usepackage[T1]{fontenc}
\usepackage{ae,aecompl}

\usepackage{graphicx}	
\usepackage{amsmath}	
\usepackage{amssymb}	
\usepackage{cancel}

\usepackage[utf8]{inputenc}
\usepackage[english]{babel}

\usepackage[usenames,dvipsnames]{xcolor}

\newcommand{\Pa} {P_\text{a}}
\newcommand{\Ps} {P_\text{s}}
\newcommand{\Pt} {P_\text{tot}}
\newcommand{\PaINI} {P_\text{a,0}}
\newcommand{\Pone} {P_\text{r}}
\newcommand{\kappaa} {\delta}

\newcommand{\gone}	{g_\text{DPL}}
\newcommand{\kone}	{k_\text{DPL}}
\newcommand{\Pdpl}	{P_\text{DPL}}
\newcommand{\Ptdpl}	{P_\text{tDPL}}
\newcommand{\Pmlp}	{P_\text{MLP}}

\title[Dual Power Law Distribution]
{A Dual Power Law Distribution for the Stellar Initial Mass Function}

\author[K.~H. Hoffmann et al.]{
Karl Heinz Hoffmann,$^{1}$
Christopher Essex,$^{2}$
Shantanu Basu$^{3}$\thanks{E-mail: basu@uwo.ca}
and Janett Prehl$^{1}$
\\
$^{1}$Institut für Physik, Technische Universität Chemnitz, D-09107 Chemnitz, Germany\\
$^{2}$Department of Applied Mathematics, University of Western Ontario, London, ON, N6A 5B7, Canada\\
$^{3}$Department of Physics \& Astronomy, University of Western Ontario, London, ON, N6A 3K7, Canada
}

\date{Accepted XXX. Received YYY; in original form ZZZ}

\pubyear{2017}

\begin{document}
\label{firstpage}
\pagerange{\pageref{firstpage}--\pageref{lastpage}}
\maketitle

\begin{abstract}
We introduce a new dual power law (DPL) probability distribution function for the mass
distribution of stellar and substellar objects at birth, otherwise known as the initial mass function (IMF). 
The model contains both deterministic and stochastic elements, and provides a unified framework
within which to view the formation of brown dwarfs and stars resulting from an accretion process that
starts from extremely low mass seeds. 
It does not depend upon a top down scenario of collapsing (Jeans) masses or
an initial lognormal or otherwise IMF-like distribution of seed masses.
Like the modified lognormal power law (MLP) distribution, the DPL distribution has a power law at the high mass end, as a result of exponential growth of mass coupled with equally likely stopping of accretion at any time interval. 
Unlike the MLP, 
a power law decay also appears at the low mass end of the IMF.
This feature is closely connected to 
the accretion stopping probability rising from an initially low value up to a high value. 
This might be associated with physical effects of ejections sometimes (i.e., rarely) stopping accretion at early times followed by outflow driven accretion stopping at later times, with the transition happening at a critical time (therefore mass).
Comparing the DPL to empirical data, the critical mass is close to 
the substellar mass limit, suggesting that the onset of nuclear fusion plays an important role in the subsequent accretion history of a young stellar object. 

\end{abstract}




\section{Introduction}

The mass distribution of stars at birth, or initial mass function (IMF), 
is an important observable characteristic of clustered star formation.
Observations show that the IMF is fairly independent of
heavy-element abundance and has a 
similar character amongst local field stars and 
in galactic and extragalactic resolved clusters 
\citep{scalo.j.86.stellar.1}. 
However, the most distinguishing characteristic of the IMF, the
high-mass tail, has measured {power law} indices that do vary within the range
$\alpha \simeq 1-2$,  
where the fractional number per logarithmic mass interval
${d}N/{d}\ln M \propto M^{-\alpha}$.
Frequently quoted values include $\alpha=1.35$ \citep{salpeter.e.55.luminosity.161}
and $\alpha=1.7$ 
\citep{kroupa.p.02.initial.82}.
Significantly,  for $\alpha \gtrsim 1${, the} vast majority of stars are found at the low mass end.
The mass content is fairly
evenly distributed, and the bulk of the total stellar luminosity emerges from the
high mass end.
The functional form of the IMF is key to studying the integrated effect 
of a collection of stars in a galaxy, 
e.g., for purposes of modeling the spectrum of a galaxy, its mass-to-light ratio,
its star formation rate, or its chemical evolution 
\citep{kennicutt.r.98.stellar.1}. 
More recent work establishes that the IMF extends into the substellar regime $M < 0.075\, M_{\odot}$ 
\citep[e.g.][]{chabrier.g.00.theory.337,%
andersen.m.08.evidence.l183} 
and also that ${d}N/{d}\ln M$ has a peak at $\sim 0.2\, M_{\odot}$ 
\citep{kroupa.p.01.variation.231,%
kroupa.p.02.initial.82,%
chabrier.g.03.galactic.763,%
chabrier.g.05.initial.41}.
The inclusion of substellar objects 
that do not achieve a steady-state nuclear fusion epoch (the main-sequence) 
makes it more important to define the ``initial'' or ``birth'' time of an object 
as the termination of mass accumulation rather than 
its appearance on the main-sequence in a Hertzsprung-Russell diagram.
It is the mass distribution of these substellar objects 
that is the focus of this paper.

One of the popular ideas for an origin of the IMF is that
turbulence in molecular clouds leads to a core mass function (CMF) that 
reflects properties of the turbulent scaling and also maps onto the 
IMF 
\citep{padoan.p.02.stellar.870,%
padoan.p.04.mysterious.559,%
hennebelle.p.08.analytical.395,%
hennebelle.p.09.analytical.1428}. 
The idea is that the CMF then 
directly maps onto the IMF since each core forms typically one or two
stars with a high, and approximately fixed, proportion of the core mass 
going into the star(s).

While turbulent fluctuations can easily produce cores with mass exceeding the local Jeans 
mass, which is the minimum mass that can collapse gravitationally,
this mechanism 
is challenged in its ability to create gravitationally collapsing low
mass cores. The collapse of a low mass core that is well below the mean 
Jeans mass requires a very coherent and focused ram pressure due to a 
turbulent flow 
\citep{lomax.o.16.forming.1242}. 
Given that the Jeans length is $\lambda_J = (\pi c_s^2/G\rho)^{1/2}$,
where
$c_s$ is the isothermal sound speed, 
$G$ is the gravitational constant,
and $\rho$ is the density, the Jeans mass is
\begin{equation}
	M_J = \rho\,\lambda_J^3 = 5.5\,M_{\odot} \left(\frac{n}{10^4\,{\rm cm}^{-3}} \right)^{-1/2} \left(\frac{T}{10\,{\rm K}} \right)^{3/2},
\end{equation}
where $n$ is the number density. It is very difficult to bring this minimum
collapsible mass down to the substellar regime, requiring an extremely high density
fluctuation. 
\citet{thies.i.15.characterizing.72} made a detailed comparison of 
observations and turbulent fragmentation models and concluded that an additional
formation channel besides the latter is required in order to explain the frequency
of substellar objects.

Years of observational surveys
of molecular clouds have also revealed only very few candidate protostellar cores 
that have substellar mass 
\citep{andre.p.12.interferometric.69,lee.c.13.early.50}.
New high sensitivity observations of young stellar clusters however, reveal an
increasing number of faint substellar objects. For example, a study of the
large massive star forming cluster RCW 38 by 
\citet{muzic.k.17.low-mass.3699} 
using adaptive optics imaging on the Very Large Telescope (VLT) reveals a
large number of spectroscopically confirmed substellar objects, 
with the ratio of 
substellar to stellar objects estimated to be about 1:2. A new survey of 
the Orion Nebula Cluster by 
\citet{drass.h.16.bimodal.1734}, 
also using the VLT, finds a ratio of
substellar to stellar objects of about 1:1, although the putative substellar
objects are not yet spectroscopically confirmed.

Altogether, the presence of abundant substellar objects in young clusters 
hints at a mechanism to produce low mass objects from parent cores that have
much greater mass. One effective way to do this is through 
disk instability in centrifugally-supported disks that form around young stars.
The high density of disks and their lifetime, that is many orbit times, allows time
for the development of gravitational instability, particularly in the early
$\lesssim 10^5$ yr of evolution, when accretion from the parent core is significant
\citep{vorobyov.e.06.burst.956,%
vorobyov.e.10.burst.1896,%
vorobyov.e.15.variable.115%
}. 
\citet{basu.s.12.hybrid.30} 
presented a scenario in which substellar mass clumps that are formed
in the disk are then ejected through multi-body interactions, leading to free-floating
proto-brown-dwarfs or proto-giant-planets. 
\citet{vorobyov.e.16.ejection.a115} 
has examined this scenario
further, and finds that ejected substellar cores can be distinguished from those
formed by direct collapse by their greater specific angular momentum.

In this paper we 
explore the possibility that the observed IMF
is set not so much by the CMF but by the accretion termination processes.
Our work is similar in concept to the IMF model of 
\cite{adams.f.96.theory.256}
in that it posits that stellar masses are set by accretion termination, 
and not
related to the Jeans mass of the cloud. 
Here, we introduce a two phase process for accretion termination, replacing the detailed
physics with a simple mathematical model.
This process is characterized by 
a small accretion termination rate in an early phase 
which then rises to a larger accretion termination rate at a later stage.

In the first phase accretion termination is occurring very rarely 
and might be due to mechanisms like disk ejection 
\citep{basu.s.12.hybrid.30,
vorobyov.e.16.ejection.a115}.
The transition to the second phase
can be associated
with the onset of nuclear (deuterium or hydrogen) fusion in the stellar
core
\citep{shu.f.87.star.23,
adams.f.96.theory.256}. 
In the second phase the accretion termination probability 
rises to a greater value when the protostellar outflows have become active. 
However, other scenarios can also qualitatively fit in this framework. 
For example, the recent simulations of 
\citet{lee.y.18.stellar.88,
lee.y.18.stellar.89} 
suggest that, in a very dense and turbulent cloud, 
accretion termination due to the formation of a neighbouring accretor 
cannot occur in gas 
that is within the tidal radius of the initial first hydrostatic core. 
The mass of gas within this tidal radius is estimated 
to be $\sim 0.1 M_{\odot}$ in their models, 
and results in a peak in their calculated IMF at about that mass, 
and independent of the large scale properties 
and mean Jeans mass of the initial model cloud.

Our model takes up features 
of the modified lognormal power law (MLP) distribution elaborated by 
\citet{basu.s.15.mlp.2413}.
The MLP is based on the idea that there is an initial lognormal
distribution of 
protostellar seeds that then grow by accretion from their surroundings at an 
exponential rate, and that there is an equally likely stopping probability for accretion
in every time interval, leading to an exponential distribution of lifetimes. 
The result is a power-law distribution of final masses at the intermediate
and high mass regime, with a power-law index that is equal to the dimensionless 
ratio of the growth time of accretion to the decay time of the exponential 
distribution of lifetimes. 

In the MLP scenario, an initial delta function
distribution of protostellar seed masses (i.e., a lognormal with zero variance)
leads to a pure power-law distribution for the IMF. However, if observations
show a low mass peak in the IMF, then the MLP model relies on a peak in the 
initial lognormal distribution of seed masses to lead to a peak in the IMF after
accretion evolution. In this paper, we explore whether a peak in the IMF can
be set by the accretion history itself, without any need for a peak in the
underlying distribution of protostellar seeds. 

Put another way, can a delta 
function of initial protostellar seed masses still yield a peaked IMF?
In this paper we explore the scenario that all protostars start their life at a 
mass much less than their final mass and undergo accretion until it is terminated. 
This initial mass is likely in the range of $\sim 10^{-3} M_{\odot}$ to $10^{-2} 
M_{\odot}$, 
based on calculations
of cloud collapse, resulting in first hydrostatic cores {that then go on to yield}
second collapse stellar seeds 
\citep{larson.r.69.numerical.271,%
masunaga.h.00.radiation.350}.
Such seeds are envisioned as the starting point of our model.

The delayed rise of the rate at which accretion is terminated
is crucial in our model. 
The resulting IMF is then naturally peaked and has power laws at both the 
low and high mass ends. The
resulting
dual power law (DPL) function differs from
lognormal or lognormal-like functions in that at the lowest masses, there will
be significantly more objects in the distribution. Distinguishing a lognormal
from a power law distribution at the low mass end of the IMF (rather than at 
the high mass end as done historically) should be an important astronomical
target in coming years as very deep measurements of young stellar clusters
are performed with new ground and space based observatories.

\section{A Model for Star Formation}

Our model captures some aspects of the random nature of the star formation process 
while also employing a deterministic accretion process. 
Here we use it to describe the evolution of mass condensations in 
gas clouds
that undergo accretion until some later time when accretion stops, 
defining the initial stellar mass.

The IMF has a probability density function denoted as $P_\text{IMF}$ 
which is a function of the mass $M$.
It is convenient to work in terms of the function $\xi=\ln m$, 
where $m \equiv M/M_\odot$ is a normalized mass.
Hereafter we use the following notation for probability density functions: 
	$P(\xi) \equiv {d}N/{d}\xi = {d}N/{d}\ln m$ 
	and $P(m) \equiv {d}N/{d}m$.
We also often refer to $\xi$ as ``mass''.

We are interested in the overall probability density $\Pt(\xi,t)$ 
to find a condensation of mass $\xi$ at time $t$, 
which obeys the normalization condition $ \int  \Pt(\xi,t)\,{d}\xi= 1 $. 
It is the sum of the two probabilities, $\Pa(\xi,t)$ and $\Ps(\xi,t)$.
$\Pa(\xi,t)$ is the probability to find a condensation of mass $\xi$ still accreting at time $t$, 
and $\Ps(\xi,t)$ is the probability to find a condensation of mass $\xi$ that 
has ceased accreting 
and has thus become a member of the IMF:
\begin{align}
\Pt(\xi,t) = \Pa(\xi,t) + \Ps(\xi,t). 
\label{eq-P}
\end{align}

The evolution of $\Pa(\xi,t)$ is governed by
\begin{align}
\partial_t \Pa(\xi,t) = {\cal L}  \Pa(\xi,t) - k(\xi,t) \Pa(\xi,t),
\label{eq-Pa-Dyn}
\end{align}
where ${\cal L}$ is an operator describing the evolution
dynamics, 
and $ k(\xi,t) $ is the \emph{accretion-dropout rate},  
with which the active condensations are becoming inactive, or stationary, 
and thus a contribution to the IMF.
The evolution operator ${\cal L}$ could be a Fokker-Planck or a master equation operator,
but could also describe just a deterministic accretion of mass. 

$\Ps(\xi,t)$ thus represents the fraction of all condensates for which accretion has ceased,
\begin{align}
\Ps(\xi,t) = \int_{t_0}^{t}  k(\xi,t') \Pa(\xi,t') {d}t',
\label{eq-Ps-Dyn}
\end{align}
where the lower integral limit $t_0$ suggests some presumed start time, 
which can be extended to $-\infty$ if desired. 
The IMF is then given by
\begin{align}
P_\text{IMF}(\xi) = \lim_{t \to \infty} \Ps(\xi,t) = \int_{t_0}^{\infty}  k(\xi,t') \Pa(\xi,t') {d}t'.
\label{eq-PIMF-Dyn}
\end{align}

The evolution equation (\ref{eq-Pa-Dyn}) is supplemented with
an initial condition $\Pa(\xi,t_0) = \PaINI(\xi)$
and boundary conditions if necessary.

We now study the implications of assumptions made in the literature
\citep{basu.s.04.power-law.L47,%
basu.s.15.mlp.2413}
 that the accretion stops with a constant rate
and that the mass accretion rate  is proportional to the already accreted mass.

\subsection{Constant Accretion-Dropout Rate}

Suppose
that the evolution operator is only dependent on $\xi$,
i.e. ${\cal L} = {\cal L}[\xi]$,
and
the rate function is set to a constant,  $ k(\xi,t) = \delta $,
then equation (\ref{eq-Pa-Dyn}) becomes
\begin{align}
\partial_t \Pa(\xi,t) ={\cal L}[\xi] \Pa(\xi,t) - \delta \Pa(\xi,t).
\label{eq-Pa-MLP-Dyn}
\end{align}

It follows from equation (\ref{eq-P}) 
that the fraction of still accreting masses,
\begin{align}
	g(t)=\int \Pa(\xi,t) {d}\xi,
\label{eq-g-Def2}
\end{align}
is smaller than unity
and thus $\Pa(\xi,t)$ is not a proper probability distribution.
But a new distribution $\Pone(\xi,t)$ can be introduced
via
$ \Pone(\xi,t) = \Pa(\xi,t) / g(t)$, 
that allows $\Pone(\xi,t)$ to integrate to unity given a particular $g(t)$.
Then $\Pone(\xi,t) \,{d}\xi$ represents that fraction of the active condensations 
that can be found in an interval ${d}\xi$ at $\xi$.

We then use  $ \Pa(\xi,t) = g(t) \Pone(\xi,t)$ as an ansatz to decompose equation (\ref{eq-Pa-MLP-Dyn}), 
\begin{align}
	\Pone(\xi,t)  \; \partial_t g(t) &+ g(t) \; \partial_t \Pone(\xi,t) 
\nonumber
\\
	&=
	g(t) {\cal L}[\xi] \Pone(\xi,t) - \delta g(t) \Pone(\xi,t).
	\label{eq-Pa-MLP-Dyn-1}
\end{align}
With the aim of conserving the probability in $\Pone(\xi,t)$, 
equation (\ref{eq-Pa-MLP-Dyn-1}) can be separated into an 
ordinary differential equation for $g(t)$ and a partial differential equation for $\Pone(\xi,t)$:
\begin{align}
\partial_t g(t) 
&=
 - \delta g(t)
\label{eq-g-MLP-Dyn} 
\\
\partial_t \Pone(\xi,t) 
&=
{\cal L}[\xi] \Pone(\xi,t).
\label{eq-Pone-MLP-Dyn2}
\end{align}
Without loss of generality 
in the following we set $t_0 = 0$.
Utilizing the initial value $g(0)=1$, 
which reflects that no condensations have dropped out at $t=0$,
we obtain
\begin{align}
g(t) &= e^{-\delta t}. 
\label{eq-g}
\end{align}
Inserting this into equation (\ref{eq-PIMF-Dyn}) then gives us the IMF as
\begin{align}
P_\text{IMF}(\xi) =  \int_{0}^{\infty}  \delta e^{-\delta t'} \Pone(\xi,t') {d}t'.
\label{eq-PIMF-Dyn-ConstantDelta}
\end{align}

\subsection{Constant Accretion Rate}

The accretion process that creates stars is a complicated phenomenon, and 
can proceed through many phases. It can resemble constant accretion at early
times 
\citep[e.g.,][]{shu.f.77.self-similar.488} 
and also exhibit episodic behavior
\citep[see][and references within]{audard.m.14.episodic.387},
but must attain greater values to create high mass stars in order to explain the
small age spread of stars of different masses in young stellar clusters 
\citep{myers.p.93.gravitational.635}.
Here we use a simplified deterministic evolution law
with the dynamics of the accretion process described by
\begin{align}
\frac{{d} m}{{d}t} = \gamma \, m,
\label{eq-LNmDot}
\end{align}
where the \emph{accretion rate} $\gamma$ gives 
the characteristic time for this process.
Based on this process 
a condensation with initial mass $m_0$ at time $t_0=0$ 
is transported in time $t$ to 
$m = m_0e^{\gamma\,t}$
and thus
the whole initial distribution is shifted to larger masses.
With the choice $\xi=\ln m$ 
the corresponding evolution operator is thus 
$ {\cal L}[\xi] = - \gamma \partial_{\xi}$
and we find
\begin{align}
\partial_t \Pone(\xi,t) 
=
 - \gamma \partial_{\xi}\Pone(\xi,t).
\label{eq-Pone-MLP-Dyn}
\end{align}
This is simply the half wave equation, 
which then propagates the initial value of $\Pa(\xi,0)$ forward in time, 
\begin{align} \Pone(\xi,t) &= \PaINI(\xi - \gamma t ).
\label{eq-Pone-MLP-DynW}
\end{align}

The MLP model 
\citep{basu.s.04.power-law.L47%
,basu.s.15.mlp.2413%
} 
has been effective in fitting the IMF, 
but it relies on a lognormal assumption to fit the low mass end,
\begin{align}
\PaINI( \ln m) \;{d} (\ln m) 
= 
\frac { \exp\left({  - \frac{ (\ln m - \mu_0)^2}  {2 \sigma_0^2}   }\right)   }   { {\sqrt{2 \pi}} \sigma_0 } \; {d} (\ln m),
\label{eq-NormalDistribution}
\end{align}
where $\mu_0$ and $\sigma_0^2$ are the mean value and the variance, 
respectively.
From here on we will use $\ln m$ instead of $\xi$ to be close
to the original form of the MLP model.

The derivation then proceeds from (\ref{eq-PIMF-Dyn}), (\ref{eq-g}), 
and (\ref{eq-Pone-MLP-Dyn}) to obtain 
\begin{align}
\Pmlp&(\ln m)  \;{d} (\ln m) 
\nonumber
\\
=& \int_{0}^{\infty}  k(\ln m,t') \Pa(\ln m,t') {d}t'  \;{d} (\ln m) 
\nonumber
\\
=& \int_{0}^{\infty}  \delta \, e^{-\delta t'} \PaINI(\ln m - \gamma t') {d}t'  \;{d} (\ln m) .
\\
=& 
\frac{\alpha}{2} \exp\left(\alpha \mu_0 + \alpha^2 \sigma_0^2/2\right)
	m ^{-\alpha} 
\nonumber
\\
& \times\,\text{erfc}\left[\frac{1}{\sqrt 2} \left(\alpha \sigma_0 - \frac{\ln m - \mu_0}{\sigma_0}\right)\right] \;{d} (\ln m) ,
\label{eq-MLP-2}
\end{align}
where $\text{erfc}$ is the complementary error function and $\alpha=\kappaa/\gamma$
is a dimensionless parameter.
But the a priori assumption of a lognormal distribution proves unnecessary 
if the dropout from accretion is simply delayed.

\section{Delayed Dropout}

As a condensate matures into a star 
it becomes more efficient at driving off infalling material 
through increasing radiation and stellar winds, 
eventually ending accretion. 

We envision two regimes: 
one where accretion is largely {not resisted}, 
and a second regime, starting at a later time through some brief transition, 
when accretion is equally likely to be stopped in any time interval. 
We allow the dropout rate to be time dependent in that it switches on at a time
later than the start of accretion.  

\subsection{Time-dependent Accretion-Dropout Rate}

For simplicity 
all condensations start with the same seed mass, $m_0$. 
Thus, 
\begin{align}
\PaINI( \ln m) \;{d} (\ln m) 
= 
\hat{\delta}(\ln m - \ln m_0) \;{d} (\ln m) ,
\label{eq-PaINI-DeltaFunc}
\end{align}
where $\hat{\delta}$ represents the Dirac delta function, 
in contrast to the unrelated constant $\delta$ used in this paper. 
The accretion process then proceeds as in the MLP
and thus (\ref{eq-Pone-MLP-Dyn}) holds,
\begin{align}
\Pone(\ln m,t) \;{d} (\ln m)
&= \hat{\delta}(\ln m- \ln m_0 - \gamma t)  \;{d} (\ln m).
\label{eq-Pone-Delta-LnM}
\end{align}

The accretion-dropout rate is chosen to be time dependent: $ k(\xi,t) = k(t) $.
With this choice, the same ansatz 
$P_a(\xi,t) = g(t)\,P_r(\xi,t)$
as above 
can be made. 
Inserting this into (\ref{eq-PIMF-Dyn}) then leads to
\begin{align}
P&_\text{IMF}(\ln m) \;{d} (\ln m)
\nonumber
\\
	&= \int_{0}^{\infty}  k(t') g(t') \hat{\delta}(\ln m- \ln m_0 - \gamma t') {d}t'  \,{d} (\ln m)
\nonumber
\\
	&= \begin{cases}
		0                                                                                &:  \;  m <m_0  \\
		\frac{ k(t_m) g(t_m)}{\gamma}   \;\;\;{d} (\ln m)   &:  \;  m \geq m_0,
	\end{cases}
\label{eq-PIMF-Delta-1}
\end{align}
where $t_m = (\ln m- \ln m_0)/{\gamma}$
is the time a condensation of mass $m$ has been accreting.

\subsection{The DPL Accretion-Dropout Rate}

The delay leads to a dual power law (DPL) for the IMF 
if the delay is characterized through the following choice for $k(t)$,
\begin{align}
\kone(t) = \kappaa (1+\tanh[(t-t_\text{S})\eta]  ),
\label{eq-k1-Def}
\end{align}
where $\kappaa$ is the 
\emph{maximum dropout rate}.
The \emph{rise time} $t_\text{S}$ indicates 
the midpoint of the transition
from smallest dropout rate to largest dropout rate. 
The \emph{dropout rise rate} $\eta$ sets the time span  $1/\eta$ of the transition zone.
Also, we note that at the dropout time one finds the greatest rate of increase of the rate function, 
$\kone(t)$.

The DPL rate is chosen such that for small times it increases slowly from a
small but non-zero starting value  
while for large times it asymptotically approaches $\delta$
after passing through a transition zone 
in which it increases rapidly.

\begin{figure}
      \includegraphics[width=0.45\textwidth]{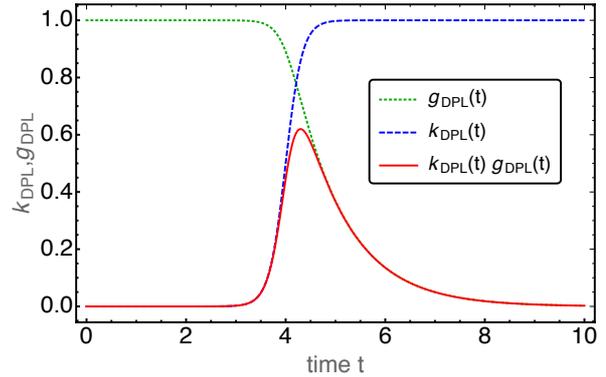}
      \caption{%
      The accretion stopping rate $\kone$ is shown as a function of the
      accretion time with $t_\text{S}=4$, $\delta=1$, and $\eta=3$.
      In addition the resulting fraction $g_\text{DPL}$ of active, i.e., 
      still accreting, condensations
      at time $t$ is depicted.
      Note that the product of the two functions shows an initial increase
      which turns into a slow exponential like decrease
      close to the crossing point of the two functions.%
      }
      \label{fig-k-g-kg}
    \end{figure}

Replacing the time independent $\delta$ in equation (\ref{eq-g-MLP-Dyn}) 
with the time dependent dropout rate $k_\text{DPL}(t)$ 
we obtain
\begin{align}
\partial_t \gone(t) 
=
 - k_\text{DPL}(t) \gone(t)
\label{eq-g-k1-Dyn}
\end{align}
with the solution
\begin{align}
\gone(t) 
=
e^{-\frac{t \kappaa}{2}}
	\left\{ \cosh[t_\text{S} \eta] /
	\cosh[(t-t_\text{S})\eta] \right\}^{\frac{\kappaa}{2 \eta}}. 
\label{eq-g-Def}
\end{align}

In Fig. \ref{fig-k-g-kg} one sees how the two functions contributing
to the IMF complement each other.
While $\kone(t)$ provides the increasing part,
$\gone(t)$ dominates the decreasing part of the IMF.

We introduce $m_\text{S}$
via $t_\text{S} = ({\ln m_\text{S}- \ln m_0})/{\gamma}$,
which is the characteristic mass 
at which the rate function has its steepest increase.
Inserting $t_\text{S}$ and $t_m$
into equation (\ref{eq-PIMF-Delta-1})
we find for $m \geq m_0$
the truncated dual power law distribution $\Ptdpl$ as
\begin{align}
&\Ptdpl(\ln m; \alpha, \beta, m_\text{S}, m_0)  \;{d} (\ln m) 
\nonumber
\\
&= \frac{ \kone(t_m) \gone(t_m)}{\gamma}  \,{d} (\ln m)
\nonumber
\\
&=\frac{\alpha}{2}  \exp\left[{-\frac{\alpha}{2}(\ln m - \ln m_0)} \right] 
\label{eq-PIMF1}
\\
	&\hspace{0.35cm}\times \{\cosh[(\ln m_0-\ln m_\text{S})\beta]/\cosh[(\ln m-\ln m_\text{S})\beta]\}^{\frac{\alpha}{2 \beta}}
\nonumber
\\ &\hspace{0.35cm}\times
			(1+\tanh[(\ln m-\ln m_\text{S})\beta])   \;\;\;{d} (\ln m),
\nonumber
\end{align}
where 
$\alpha=\kappaa/\gamma$ 
as above and
$\beta=\eta/\gamma$
is a further dimensionless parameter.

\subsection{The DPL Distribution}

A careful inspection of equation (\ref{eq-PIMF1}) reveals 
that 
$\Ptdpl$
converges rapidly
towards a limiting distribution for $m_0 \to 0$,
the dual power law (DPL) distribution:
\begin{align}
&\Pdpl(m; \alpha, \beta, m_\text{S}) \;{d}m
=
\frac{\alpha}{m}\,
2^{  -\frac{\alpha}{2 \beta}-1  }
\exp\left[{-\frac{\alpha}{2} \ln\frac{m}{m_\text{S}}} \right]
\nonumber
\\
&\hspace{0.15cm}
\times \left\{\cosh\left[\beta\,\ln\frac{m}{m_\text{S}}\right]
	\right\}^{-\frac{\alpha}{2 \beta}}   		
\,\left(1+\tanh\left[\beta\,\ln\frac{m}{m_\text{S}}\right]\right)   \;\;{d}m \,.
\label{eq-DPL-m}
\end{align}
The analysis of the DPL distribution also shows 
that its decay towards smaller and larger masses 
are described by power laws $\Pdpl(m) \propto m^{\nu}$.
The exponents $\nu_{m > m_\text{S}}$ and $\nu_{m < m_\text{S}}$
of those power law tails are related to the
model parameters in a simple fashion:
$\nu_{m > m_\text{S}} = -(\alpha + 1)$
and
$\nu_{m<m_\text{S}} = 2\,\beta-1$.

The surprising and unexpected feature of the DPL function
based on the model assumptions made
is the power law behavior of the IMF at the low mass tail.
Such a power law behavior of the IMF at the low mass end
has been suggested in the literature before.
\citet{rio.n.12.initial.14} 
performed extensive optical observations of the Orion Nebula Cluster (ONC), 
and compiled the masses of confirmed cluster members.
Their deduced IMF could be fitted by power law functions on both sides 
of a peak value,
when using one of the two considered evolutionary models.

\begin{figure}
      \includegraphics[width=0.45\textwidth]{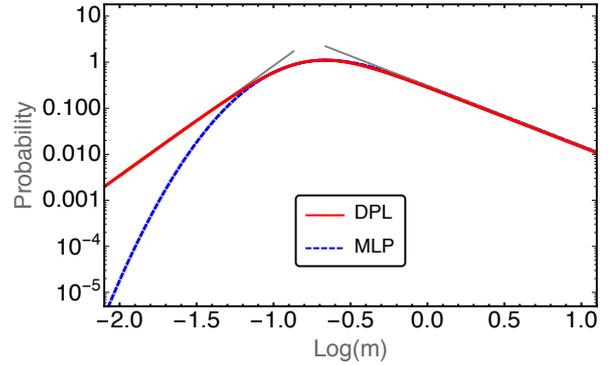}
      \caption{
		The DPL distribution is shown together with the MLP. 
		The gray lines indicate the asymtotic power law tails
		of the DPL at the high and low mass ends.
		The parameters are chosen such that DPL and MLP
		have the same behavior at the high mass end
		and show a clear difference at the low mass end of the IMF.
		 }
      \label{fig-DPL-MLP}
    \end{figure}

In Fig.~\ref{fig-DPL-MLP} we plot 
the DPL function that utilizes the same values of 
$\alpha (=1.3)$ and $\beta (=1.205)$ 
as the fit by 
\citet{rio.n.12.initial.14} 
(see their table 4). 
The asymtotic power law behavior is indicated by gray lines
on both sides of the peak. 
The DPL is a continuous function and is shown here 
for $m_\text{S}=10^{-0.78}=0.166$. 
In addition we plot a MLP with same high mass power exponent 
$\alpha =1.3$,
$\mu_0= -2$, and $\sigma_0= 0.6$.
One clearly sees the difference between the DPL and the MLP at the
low mass end.

Interestingly the value of $m_\text{S}$
that is compatible with the 
\citet{rio.n.12.initial.14} data
is about a factor of 2 
within the substellar mass limit that has a typical value
of $0.075\, M_{\odot}$
\citep{chabrier.g.00.theory.337} 
and an estimated range of uncertainty $0.064\, M_{\odot} - 0.087\, M_{\odot}$  
\citep{auddy.s.16.analytic.5743272}.
We also note  
that $m_\text{S}$ is not the mass at which 
the transition begins, but rather the mass at which a steepest transition in the 
accretion stopping rate is taking place.
This raises the possibility of the physics behind the transition 
being related to the onset of nuclear fusion. 
While this is a tempting possibility, 
this question cannot be answered within the
context of the mathematical model
presented here.

The importance of a DPL function for the IMF, 
if verified by future observations, 
is two-fold. 
It implies the existence of a generative effect for the IMF 
that operates at low masses, 
and is not dependent on an imprinted lognormal distribution
from stochastic processes. 
We have presumed that the generative effect is accretion, 
and the lack of accretion stopping at very early times 
leads to the power law at low masses. 
Secondly, if the very low mass IMF is a power law, 
one can reasonably expect that below some small enough value of $m$,
there will be more objects than predicted by a lognormal decay, 
no matter what is
the value of the power law index.

\section{Conclusions}

Our model uses a simplified scenario for stellar mass accretion 
and delayed accretion dropout 
in order to derive an analytic three parameter dual power law (DPL) probability distribution function 
for the IMF. 
The two power laws have different exponents. 
Each has a direct physical interpretation.

The high mass tail has an exponent set by the ratio
$\alpha=\kappaa/\gamma$ of 
the maximum dropout rate $\kappaa$ and 
the accretion rate $\gamma$ 
as in the MLP model. 
The low mass tail has an exponent that depends on the ratio 
$\beta=\eta/\gamma$
of 
the dropout rise rate $\eta$ and 
the accretion rate $\gamma$.
The third dimensionless parameter of the distribution is related to the 
characteristic mass $m_\text{S}$ at which the dropout rate rises most
rapidly. 

A remarkable feature of the empirical IMF is 
that this characteristic mass is so similar to the substellar mass limit. 
As the empirical IMF of different star clusters vary considerably 
\citep{dib.s.14.testing.1957},
it would be premature to predict that this feature will occur in all of them.
Nonetheless, this proximity raises the question of
whether the peak of the IMF can be traced to a transition 
originating from fundamental physics, 
specifically the onset of
nuclear fusion.

Our model provides insight into 
a possible scenario of generative processes 
for a dual power law IMF, 
but does not require 
that identical values of the power-law indices 
are present in each star-forming cluster, 
as these may depend on the specific accretion 
and dropout histories in that cluster. 
What we emphasize is that
a dual power law distribution is qualitatively different from 
a lognormal-like decrease, and has
a larger number of substellar objects below a certain mass. 
If the dual power law is realized in nature, then 
there could be a large population of
currently undetected very low mass substellar objects in the Galaxy.


\

\section*{Acknowledgements}
SB was supported by an NSERC grant. 
Thanks go to Deepakshi Madaan
for her input.
We also thank the referee for helpful comments.




\bibliographystyle{mnras}
\bibliography{ms_final}

\begin{thebibliography}{}
\makeatletter
\relax
\def\mn@urlcharsother{\let\do\@makeother \do\$\do\&\do\#\do\^\do\_\do\%\do\~}
\def\mn@doi{\begingroup\mn@urlcharsother \@ifnextchar [ {\mn@doi@}
  {\mn@doi@[]}}
\def\mn@doi@[#1]#2{\def\@tempa{#1}\ifx\@tempa\@empty \href
  {http://dx.doi.org/#2} {doi:#2}\else \href {http://dx.doi.org/#2} {#1}\fi
  \endgroup}
\def\mn@eprint#1#2{\mn@eprint@#1:#2::\@nil}
\def\mn@eprint@arXiv#1{\href {http://arxiv.org/abs/#1} {{\tt arXiv:#1}}}
\def\mn@eprint@dblp#1{\href {http://dblp.uni-trier.de/rec/bibtex/#1.xml}
  {dblp:#1}}
\def\mn@eprint@#1:#2:#3:#4\@nil{\def\@tempa {#1}\def\@tempb {#2}\def\@tempc
  {#3}\ifx \@tempc \@empty \let \@tempc \@tempb \let \@tempb \@tempa \fi \ifx
  \@tempb \@empty \def\@tempb {arXiv}\fi \@ifundefined
  {mn@eprint@\@tempb}{\@tempb:\@tempc}{\expandafter \expandafter \csname
  mn@eprint@\@tempb\endcsname \expandafter{\@tempc}}}

\bibitem[\protect\citeauthoryear{Adams \& Fatuzzo}{Adams \&
  Fatuzzo}{1996}]{adams.f.96.theory.256}
Adams F.~C.,  Fatuzzo M.,  1996, \mn@doi [Astrophys. J.] {10.1086/177318}, 464,
  256

\bibitem[\protect\citeauthoryear{Andersen, Meyer, Greissl  \& Aversa}{Andersen
  et~al.}{2008}]{andersen.m.08.evidence.l183}
Andersen M.,  Meyer M.~R.,  Greissl J.,   Aversa A.,  2008, \mn@doi [Astrophys.
  J. Lett.] {10.1086/591473}, 683, L183

\bibitem[\protect\citeauthoryear{Andr{\'e}, Ward-Thompson  \&
  Greaves}{Andr{\'e} et~al.}{2012}]{andre.p.12.interferometric.69}
Andr{\'e} P.,  Ward-Thompson D.,   Greaves J.,  2012, \mn@doi [Science]
  {10.1126/science.1222602}, 337, 69

\bibitem[\protect\citeauthoryear{Audard et~al.,}{Audard
  et~al.}{2014}]{audard.m.14.episodic.387}
Audard M.,  et~al., 2014, in Beuther H.,  Klessen R.~S.,  Dullemond C.~P.,
  Henning T.,  eds, , Protostars and Planets VI.
University of Arizona Press, Tuscon, pp 387--410,
  \mn@doi{10.2458/azu\_uapress\_9780816531240-ch017}

\bibitem[\protect\citeauthoryear{Auddy, Basu  \& Valluri}{Auddy
  et~al.}{2016}]{auddy.s.16.analytic.5743272}
Auddy S.,  Basu S.,   Valluri S.~R.,  2016, \mn@doi [Adv. Astron.]
  {10.1155/2016/5743272}, 2016, 5743272

\bibitem[\protect\citeauthoryear{Basu \& Jones}{Basu \&
  Jones}{2004}]{basu.s.04.power-law.L47}
Basu S.,  Jones C.~E.,  2004, \mn@doi [Mon. Not. R. Astron. Soc.]
  {10.1111/j.1365-2966.2004.07405.x}, 347, L47

\bibitem[\protect\citeauthoryear{Basu \& Vorobyov}{Basu \&
  Vorobyov}{2012}]{basu.s.12.hybrid.30}
Basu S.,  Vorobyov E.~I.,  2012, \mn@doi [Astrophys. J.]
  {10.1088/0004-637X/750/1/30}, 750, 30

\bibitem[\protect\citeauthoryear{Basu, Gil  \& Auddy}{Basu
  et~al.}{2015}]{basu.s.15.mlp.2413}
Basu S.,  Gil M.,   Auddy S.,  2015, \mn@doi [Mon. Not. R. Astron. Soc.]
  {10.1093/mnras/stv445}, 449, 2413

\bibitem[\protect\citeauthoryear{Chabrier}{Chabrier}{2003}]{chabrier.g.03.galactic.763}
Chabrier G.,  2003, \mn@doi [Publ. Astron. Soc. Pac.] {10.1086/376392}, 115,
  763

\bibitem[\protect\citeauthoryear{Chabrier}{Chabrier}{2005}]{chabrier.g.05.initial.41}
Chabrier G.,  2005, in Corbelli E.,  Palla F.,   Zinnecker H.,  eds,
  Astrophysics and Space Science Library Vol. 327, The Initial Mass Function 50
  Years Later. Springer Netherlands, pp 41--50,
  \mn@doi{10.1007/978-1-4020-3407-7\_5}

\bibitem[\protect\citeauthoryear{Chabrier \& Baraffe}{Chabrier \&
  Baraffe}{2000}]{chabrier.g.00.theory.337}
Chabrier G.,  Baraffe I.,  2000, \mn@doi [Annu. Rev. Astron. Astrophys.]
  {10.1146/annurev.astro.38.1.337}, 38, 337

\bibitem[\protect\citeauthoryear{Da~Rio, Robberto, Hillenbrand, Henning  \&
  Stassun}{Da~Rio et~al.}{2012}]{rio.n.12.initial.14}
Da~Rio N.,  Robberto M.,  Hillenbrand L.~A.,  Henning T.,   Stassun K.~G.,
  2012, \mn@doi [Astro. J.] {10.1088/0004-637X/748/1/14}, 748, 14

\bibitem[\protect\citeauthoryear{Dib}{Dib}{2014}]{dib.s.14.testing.1957}
Dib S.,  2014, \mn@doi [Mon. Not. R. Astron. Soc.] {10.1093/mnras/stu1521},
  444, 1957

\bibitem[\protect\citeauthoryear{Drass, Haas, Chini, Bayo, Hackstein,
  Hoffmeister, Godoy  \& Vogt}{Drass et~al.}{2016}]{drass.h.16.bimodal.1734}
Drass H.,  Haas M.,  Chini R.,  Bayo A.,  Hackstein M.,  Hoffmeister V.,  Godoy
  N.,   Vogt N.,  2016, \mn@doi [Mon. Not. R. Astron. Soc.]
  {10.1093/mnras/stw1094}, 461, 1734

\bibitem[\protect\citeauthoryear{Hennebelle \& Chabrier}{Hennebelle \&
  Chabrier}{2008}]{hennebelle.p.08.analytical.395}
Hennebelle P.,  Chabrier G.,  2008, Astrophys. J., 684, 395

\bibitem[\protect\citeauthoryear{Hennebelle \& Chabrier}{Hennebelle \&
  Chabrier}{2009}]{hennebelle.p.09.analytical.1428}
Hennebelle P.,  Chabrier G.,  2009, \mn@doi [Astrophys. J.]
  {10.1088/0004-637X/702/2/1428}, 702, 1428

\bibitem[\protect\citeauthoryear{Kennicutt}{Kennicutt}{1998}]{kennicutt.r.98.stellar.1}
Kennicutt R. C.~J.,  1998, in Gilmore G.,  Howell D.,  eds,  Astronomical
  Society of the Pacific Conference Series Vol. 142, The Stellar Initial Mass
  Function (38th Herstmonceux Conference). Astronomical Society of the Pacific,
  pp 1--15

\bibitem[\protect\citeauthoryear{Kroupa}{Kroupa}{2001}]{kroupa.p.01.variation.231}
Kroupa P.,  2001, \mn@doi [Mon. Not. R. Astron. Soc.]
  {10.1046/j.1365-8711.2001.04022.x}, 322, 231

\bibitem[\protect\citeauthoryear{Kroupa}{Kroupa}{2002}]{kroupa.p.02.initial.82}
Kroupa P.,  2002, \mn@doi [Science] {10.1126/science.1067524}, 295, 82

\bibitem[\protect\citeauthoryear{Larson}{Larson}{1969}]{larson.r.69.numerical.271}
Larson R.~B.,  1969, \mn@doi [Mon. Not. R. Astron. Soc.]
  {10.1093/mnras/145.3.271}, 145, 271

\bibitem[\protect\citeauthoryear{{Lee} \& {Hennebelle}}{{Lee} \&
  {Hennebelle}}{2018a}]{lee.y.18.stellar.88}
{Lee} Y.-N.,  {Hennebelle} P.,  2018a, \mn@doi [\aap]
  {10.1051/0004-6361/201731522}, \href
  {http://adsabs.harvard.edu/abs/2018A%26A...611A..88L} {611, A88}

\bibitem[\protect\citeauthoryear{{Lee} \& {Hennebelle}}{{Lee} \&
  {Hennebelle}}{2018b}]{lee.y.18.stellar.89}
{Lee} Y.-N.,  {Hennebelle} P.,  2018b, \mn@doi [\aap]
  {10.1051/0004-6361/201731523}, \href
  {http://adsabs.harvard.edu/abs/2018A%26A...611A..89L} {611, A89}

\bibitem[\protect\citeauthoryear{Lee, Kim, Kim, Saito, Myers  \& Kurono}{Lee
  et~al.}{2013}]{lee.c.13.early.50}
Lee C.~W.,  Kim M.-R.,  Kim G.,  Saito M.,  Myers P.~C.,   Kurono Y.,  2013,
  \mn@doi [Astrophys. J.] {10.1088/0004-637X/777/1/50}, 777, 50

\bibitem[\protect\citeauthoryear{Lomax, Whitworth  \& Hubber}{Lomax
  et~al.}{2016}]{lomax.o.16.forming.1242}
Lomax O.,  Whitworth A.~P.,   Hubber D.~A.,  2016, \mn@doi [Mon. Not. R.
  Astron. Soc.] {10.1093/mnras/stw406}, 458, 1242

\bibitem[\protect\citeauthoryear{Masunaga \& Inutsuka}{Masunaga \&
  Inutsuka}{2000}]{masunaga.h.00.radiation.350}
Masunaga H.,  Inutsuka S.-I.,  2000, Astrophys. J., 531, 350

\bibitem[\protect\citeauthoryear{Mu\v{z}i\'c, Sch\"odel, Scholz, Geers,
  Jayawardhana, Ascenso  \& Cieza}{Mu\v{z}i\'c
  et~al.}{2017}]{muzic.k.17.low-mass.3699}
Mu\v{z}i\'c K.,  Sch\"odel R.,  Scholz A.,  Geers V.~C.,  Jayawardhana R.,
  Ascenso J.,   Cieza L.~A.,  2017, \mn@doi [Mon. Not. R. Astron. Soc.]
  {10.1093/mnras/stx1906}, 471, 3699

\bibitem[\protect\citeauthoryear{Myers \& Fuller}{Myers \&
  Fuller}{1993}]{myers.p.93.gravitational.635}
Myers P.~C.,  Fuller G.~A.,  1993, \mn@doi [Astrophys. J.] {10.1086/172165},
  402, 635

\bibitem[\protect\citeauthoryear{Padoan \& Nordlund}{Padoan \&
  Nordlund}{2002}]{padoan.p.02.stellar.870}
Padoan P.,  Nordlund {\AA}.,  2002, Astrophys. J., 576, 870

\bibitem[\protect\citeauthoryear{Padoan \& Nordlung}{Padoan \&
  Nordlung}{2004}]{padoan.p.04.mysterious.559}
Padoan P.,  Nordlung {\AA}.,  2004, Astrophys. J., 617, 559

\bibitem[\protect\citeauthoryear{Salpeter}{Salpeter}{1955}]{salpeter.e.55.luminosity.161}
Salpeter E.~E.,  1955, \mn@doi [Astrophys. J.] {10.1086/145971}, 121, 161

\bibitem[\protect\citeauthoryear{Scalo}{Scalo}{1986}]{scalo.j.86.stellar.1}
Scalo J.~M.,  1986, Fundamentals of Cosmic Physics, 11, 1

\bibitem[\protect\citeauthoryear{Shu}{Shu}{1977}]{shu.f.77.self-similar.488}
Shu F.~H.,  1977, \mn@doi [Astrophys. J.] {10.1086/155274}, 214, 488

\bibitem[\protect\citeauthoryear{Shu, Adams  \& Lizano}{Shu
  et~al.}{1987}]{shu.f.87.star.23}
Shu F.~H.,  Adams F.~C.,   Lizano S.,  1987, \mn@doi [Annu. Rev. Astron.
  Astrophys.] {10.1146/annurev.aa.25.090187.000323}, 25, 23

\bibitem[\protect\citeauthoryear{Thies, Pflamm-Altenburg, Kroupa  \&
  Marks}{Thies et~al.}{2015}]{thies.i.15.characterizing.72}
Thies I.,  Pflamm-Altenburg J.,  Kroupa P.,   Marks M.,  2015, \mn@doi
  [Astrophys. J.] {10.1088/0004-637X/800/1/72}, 800, 72

\bibitem[\protect\citeauthoryear{Vorobyov}{Vorobyov}{2016}]{vorobyov.e.16.ejection.a115}
Vorobyov E.~I.,  2016, \mn@doi [Astron. Astrophys.]
  {10.1051/0004-6361/201628102}, 590, A115

\bibitem[\protect\citeauthoryear{Vorobyov \& Basu}{Vorobyov \&
  Basu}{2006}]{vorobyov.e.06.burst.956}
Vorobyov E.~I.,  Basu S.,  2006, \mn@doi [Astrophys. J.] {10.1086/507320}, 650,
  956

\bibitem[\protect\citeauthoryear{Vorobyov \& Basu}{Vorobyov \&
  Basu}{2010}]{vorobyov.e.10.burst.1896}
Vorobyov E.~I.,  Basu S.,  2010, \mn@doi [Astrophys. J.]
  {10.1088/0004-637X/719/2/1896}, 719, 1896

\bibitem[\protect\citeauthoryear{Vorobyov \& Basu}{Vorobyov \&
  Basu}{2015}]{vorobyov.e.15.variable.115}
Vorobyov E.~I.,  Basu S.,  2015, \mn@doi [Astrophys. J.]
  {10.1088/0004-637X/805/2/115}, 805, 115

\makeatother
\end{thebibliography}



%
%
%

\bsp	
\label{lastpage}
\end{document}